\input harvmac.tex


\def\sp{\partial\!\!\!\!/}
\def\sv{\varphi\!\!\!\!/}
\def\sf{F\!\!\!\!/}

\def\snabla{ \nabla \!\!\!\!/}

\lref\Metsaev{ R.~R.~Metsaev, ``Type IIB Green-Schwarz superstring in plane wave Ramond-Ramond  background,''
Nucl.\ Phys.\ B {\bf 625}, 70 (2002) [arXiv:hep-th/0112044].
 ; R.~R.~Metsaev and A.~A.~Tseytlin, ``Exactly solvable model of superstring in plane wave
Ramond-Ramond  background,'' arXiv:hep-th/0202109}

\lref\LG{K.~Hori and C.~Vafa, ``Mirror symmetry,'' arXiv:hep-th/0002222}

\lref\WitBag{J.~Bagger and E.~Witten, ``The Gauge Invariant Supersymmetric Nonlinear Sigma Model,'' Phys.\ Lett.\
B {\bf 118}, 103 (1982).}

\lref\Freed{L.~Alvarez-Gaume and D.~Z.~Freedman, ``Potentials For The Supersymmetric Nonlinear Sigma Model,''
Commun.\ Math.\ Phys.\  {\bf 91}, 87 (1983)}

\lref\Gates{S.~J.~Gates, ``Superspace Formulation Of New Nonlinear Sigma Models,'' Nucl.\ Phys.\ B {\bf 238}, 349
(1984)}

\lref\BMN{D.~Berenstein, J.~M.~Maldacena and H.~Nastase, ``Strings in flat space and pp waves from N = 4 super
Yang Mills,'' JHEP {\bf 0204}, 013 (2002) [arXiv:hep-th/0202021]}

\lref\GSW{ M.~B.~Green, J.~H.~Schwarz, E.~Witten, ``Superstring Theory,'' vol. II, Cambridge 1987, pp. 440}

\lref\Blau{M.~Blau, J.~Figueroa-O'Farrill, C.~Hull and G.~Papadopoulos, ``A new maximally supersymmetric
background of IIB superstring theory,'' JHEP {\bf 0201}, 047 (2002) [arXiv:hep-th/0110242]}

\lref\Cvetic{M.~Cvetic, H.~Lu and C.~N.~Pope, ``M-theory pp-waves, Penrose limits and supernumerary
supersymmetries,'' arXiv:hep-th/0203229}

\lref\EHorig{T.~Eguchi, P.~B.~Gilkey and A.~J.~Hanson, ``Gravitation, Gauge Theories And Differential Geometry,''
Phys.\ Rept.\  {\bf 66}, 213 (1980)}

\lref\EHkahler{A.~Hatzinikitas, ``Classical and quantum motion on the orbifold limit of the Eguchi-Hanson
metric,'' Nuovo Cim.\ B {\bf 114}, 427 (1999) [arXiv:hep-th/9808015]}

\lref\CYeight{E.~G.~Floratos and A.~Kehagias, ``Eight-dimensional self-dual spaces,'' Phys.\ Lett.\ B {\bf 427},
283 (1998) [arXiv:hep-th/9802107]}

\lref\Floratos{ E.~Floratos and A.~Kehagias, ``Penrose limits of orbifolds and orientifolds,''
arXiv:hep-th/0203134}

\lref\Gseven{S.~L.~Shatashvili and C.~Vafa, ``Superstrings And Manifold Of Exceptional Holonomy,''
arXiv:hep-th/9407025}

\lref\Ahn{C.~R.~Ahn, ``Complete S Matrices Of Supersymmetric Sine-Gordon Theory And Perturbed Superconformal
Minimal Model,'' Nucl.\ Phys.\ B {\bf 354}, 57 (1991)}

\lref\Ahnb{C.~Ahn, D.~Bernard and A.~LeClair, ``Fractional Supersymmetries In Perturbed Coset Cfts And Integrable
Soliton Theory,'' Nucl.\ Phys.\ B {\bf 346}, 409 (1990).}

\lref\Kob {K.~I.~Kobayashi and T.~Uematsu, ``S matrix of N=2 supersymmetric Sine-Gordon theory,'' Phys.\ Lett.\ B
{\bf 275}, 361 (1992) [arXiv:hep-th/9110040]}

\lref\Witten{ R.~Shankar and E.~Witten, ``The S Matrix Of The Supersymmetric Nonlinear Sigma Model,'' Phys.\ Rev.\
D {\bf 17}, 2134 (1978).}

\lref\Schoutens{ K.~Schoutens, ``Supersymmetry And Factorizable Scattering,'' Nucl.\ Phys.\ B {\bf 344}, 665
(1990)}

\lref\zam{A.~B.~Zamolodchikov and A.~B.~Zamolodchikov, ``Factorized S-Matrices In Two Dimensions As The Exact
Solutions Of  Certain Relativistic Quantum Field Models,'' Annals Phys.\  {\bf 120}, 253 (1979)}

\lref\zamsausage{
V.~A.~Fateev, E.~Onofri and A.~B.~Zamolodchikov,
``The Sausage model (integrable deformations of O(3) sigma model),''
Nucl.\ Phys.\ B {\bf 406}, 521 (1993).
}

\lref\zamlu{
V.~V.~Bazhanov, S.~L.~Lukyanov and A.~B.~Zamolodchikov,
``Quantum field theories in finite volume: Excited state energies,''
Nucl.\ Phys.\ B {\bf 489}, 487 (1997)
[arXiv:hep-th/9607099];
P.~Dorey and R.~Tateo,
``Excited states by analytic continuation of TBA equations,''
Nucl.\ Phys.\ B {\bf 482}, 639 (1996)
[arXiv:hep-th/9607167].
}

\lref\Thun{M.~Karowski and H.~J.~Thun, ``Complete S Matrix Of The Massive Thirring Model,'' Nucl.\ Phys.\ B {\bf
130}, 295 (1977)}

\lref\Fendley{P.~Fendley and K.~A.~Intriligator, ``Scattering and thermodynamics of fractionally charged
supersymmetric solitons,'' Nucl.\ Phys.\ B {\bf 372}, 533 (1992) [arXiv:hep-th/9111014]}

\lref\FendleyB{P.~Fendley and K.~A.~Intriligator, ``Scattering and thermodynamics in integrable N=2 theories,''
Nucl.\ Phys.\ B {\bf 380}, 265 (1992) [arXiv:hep-th/9202011]. }

\lref\Toda{ P.~Fendley, W.~Lerche, S.~D.~Mathur and N.~P.~Warner, ``N=2 Supersymmetric Integrable Models From
Affine Toda Theories,'' Nucl.\ Phys.\ B {\bf 348}, 66 (1991) }

\lref\Todab{E.~Ogievetsky and P.~Wiegmann, ``Factorized S Matrix And The Bethe Ansatz For Simple Lie Groups,''
Phys.\ Lett.\ B {\bf 168}, 360 (1986). }

\lref\Lvl{ C.~B.~Thorn, ``The S Matrix And The Effective Potential For The Liouville Quantum Field Theory,''
Phys.\ Lett.\ B {\bf 128}, 207 (1983). }

\lref\Bethe{ H.~Bethe, ``On The Theory of Metals. 1. Eigenvalues And Eigenfunctions For The Linear Atomic Chain,''
Z.\ Phys.\ {\bf 71}, 205 (1931). ; C.~N.~Yang and C.~P.~Yang, ``Thermodynamics of One-Dimensional System of Bosons
With Repulsive Delta Function Interaction,'' J.\ Math\. Phys.\ {\bf 10}, 1115 (1969). ; A.~B.~Zamolodchikov,
``Thermodynamic Bethe Ansatz In Relativistic Models. Scaling Three States Potts and Lee-Yang Models,'' Nucl.\
Phys.\ B {\bf 342}, 695 (1990).}

\lref\Olalla{ O.~A.~ Castro-Alvaredo, ``Bootstrap Methods in 1+1 dimensional quantum field theories: The
homogeneous sine-Gordon models,'' arXiv:hep-th/0109212.}

\lref\Miramontes{ C.~R.~Fernandez-Pousa, M.~V.~Gallas, T.~J.~Hollowood and J.~L.~Miramontes, ``Solitonic
integrable perturbations of parafermionic theories,'' Nucl.\ Phys.\ B {\bf 499}, 673 (1997).}

\lref\Rudd{R.~E.~Rudd, ``Light cone gauge quantization of 2-D sigma models,'' Nucl.\ Phys.\ B {\bf 427}, 81 (1994)
[arXiv:hep-th/9402106].}

\lref\fintr{ P. Fendley and K. Intrilligator, unpublished}

\lref\hk{ K. Hori and A. Kapustin
``Duality of the fermionic 2d black hole and N = 2 Liouville theory as  mirror symmetry,''
JHEP {\bf 0108}, 045 (2001)
[arXiv:hep-th/0104202].}

\lref\martinec{
E.~J.~Martinec and W.~McElgin,
``Exciting AdS orbifolds,''
arXiv:hep-th/0206175.
}

\lref\berkovits{N.~Berkovits,
``The Heterotic Green-Schwarz superstring on an N=(2,0) superworldsheet,''
Nucl.\ Phys.\ B {\bf 379}, 96 (1992)
[arXiv:hep-th/9201004];
``Calculation of Green-Schwarz superstring amplitudes using the N=2 twistor string formalism,''
Nucl.\ Phys.\ B {\bf 395}, 77 (1993)
[arXiv:hep-th/9208035].
}

\lref\nbjm{
N. Berkovits and J. Maldacena, to appear.}

\lref\kkk{
V.~Kazakov, I.~K.~Kostov and D.~Kutasov,
``A matrix model for the two-dimensional black hole,''
Nucl.\ Phys.\ B {\bf 622}, 141 (2002)
[arXiv:hep-th/0101011].
}
\lref\jor{
A.~N.~Jourjine,
``The Effective Potential In Extended Supersymmetric Nonlinear Sigma Models,''
Annals Phys.\  {\bf 157}, 489 (1984);
A.~N.~Jourjine,
``Constraints On Superpotentials In Off-Shell Extended Nonlinear Sigma Models,''
Nucl.\ Phys.\ B {\bf 236}, 181 (1984).
}

\lref\wbag{J.~Bagger and J.~Wess,
``Supersymmetry And Supergravity,'' Princeton Univ. Press, 1992.}

\lref\hvopen{
K.~Hori, A.~Iqbal and C.~Vafa,
``D-branes and mirror symmetry,''
arXiv:hep-th/0005247.
}

\lref\tseytlinmass{
A.~A.~Tseytlin,
``Finite sigma models and exact string solutions with Minkowski signature metric,''
Phys.\ Rev.\ D {\bf 47}, 3421 (1993)
[arXiv:hep-th/9211061];
``String vacuum backgrounds with covariantly constant null Killing vector and 2-d quantum gravity,''
Nucl.\ Phys.\ B {\bf 390}, 153 (1993)
[arXiv:hep-th/9209023];
``A Class of finite two-dimensional sigma models and string vacua,''
Phys.\ Lett.\ B {\bf 288}, 279 (1992)
[arXiv:hep-th/9205058].
}
\lref\klim{
C.~Klimcik and A.~A.~Tseytlin,
Nucl.\ Phys.\ B {\bf 424}, 71 (1994)
[arXiv:hep-th/9402120].
}

{ \Title{\vbox{\baselineskip12pt
\hbox{hep-th/0207284}
}}
{\vbox{
{\centerline { Strings on pp-waves and }}
{\centerline {massive two dimensional field theories }}
  }} }
\bigskip
\centerline{ Juan Maldacena$^{1}$ and
Liat Maoz$^{2,3}$  }
\bigskip
\centerline{$^1$ Institute for Advanced Study}
\centerline{Princeton, NJ 08540,USA}
\bigskip
\centerline{$^2$ Jefferson Physical Laboratory}
\centerline{Harvard University}
\centerline{Cambridge, MA 02138, USA}
\bigskip
\centerline{$^3$ Jadwin Hall}
\centerline{Princeton University}
\centerline{Princeton, NJ 08544,USA}

\vskip .3in

We find a general class of pp-wave solutions of type IIB string
theory  such that the light cone gauge worldsheet
lagrangian is that of an interacting massive field theory.
When the light cone Lagrangian has (2,2) supersymmetry we can
find backgrounds that lead to
arbitrary superpotentials on the worldsheet.
We consider situations with both flat and curved transverse spaces.
We describe in some detail the background giving rise to the $N=2$ sine
Gordon theory on the worldsheet.
Massive mirror symmetry relates it to the deformed
$CP^1$ model (or sausage model) which seems to elude a purely supergravity
target space interpretation.


\newsec{Introduction}

Ramond-Ramond backgrounds are a very important piece of string theory and they play a prominent role in the string
theory/gauge theory correspondence. Backgrounds of the plane wave type are particularly interesting since they are
the only known exactly solvable backgrounds \Metsaev .
 These backgrounds are very useful for studying the relation between
large $N$ gauge theory and string theory \BMN .
 The existence of a covariantly constant null Killing vector  greatly
simplifies the quantization of a string in light cone gauge. In this paper we study backgrounds of the pp-wave
type which
lead to {\it interacting} theories in light cone gauge. For this purpose we consider type IIB string theory with a
five-form field strength which has the form $F_5 = dx^+ \wedge \varphi_4$. If $\varphi_4$ is a constant form in
the transverse space it leads to masses for the Green-Schwarz light cone fermions. By taking non-constant four
forms $\varphi_4$ we find that the light cone action becomes an interacting theory with a rather general
potential. The mass scale in the light cone theory is set by $p_-$. Boosts in the $x^+, ~x^-$ directions corresponds
to an RG flow transformation  on the worldsheet. Low values of $|p_-|$ correspond to the UV of the worldsheet theory
while large values of $|p_-|$ explore the IR of the worldsheet theory.
We  study solutions that preserve some
supersymmetries. We  find that we can have an $N=(2,2) $ theory on the worldsheet with an arbitrary
superpotential. Similarly we can get $N=(1,1)$ theories as long as the real superpotential is a harmonic function.
We  discuss solutions where the transverse space is curved or flat. One interesting result is that we can find
backgrounds that lead to integrable models on the worldsheet in light cone gauge. Using results for integrable
models we can compute some non-trivial features of the string spectrum. We can consider for example Toda theories.
We discuss explicitly the case where we get the $N=2$ sine Gordon model on the worldsheet. Soliton solutions of
the massive theory correspond to strings that interpolate between different ``potential wells'' in the target
space. Now that we have massive interacting theories on the worldsheet we see that various dualities of these
theories are worldsheet dualities which lead to interesting dualities in the target space. The $N=2$ sine Gordon
theory is dual to the supersymmetric $CP^1$ theory \refs{\Fendley,\FendleyB,\fintr,\LG,\hk}, via a
mirror symmetry transformation. The size of
the $CP^1$ depends on the energy scale of the worldsheet theory. The size of the worldsheet circle is proportional
to $p_-$. Thus, we  find that strings with very small $p_-$ feel  they are on a big space while strings with large
$p_-$ feel they are on a smaller space.

Other backgrounds that lead to interacting theories in lightcone gauge
were described in \refs{\tseytlinmass,\klim}.

In section two we discuss the gravity backgrounds that lead to supersymmetric interacting theories on the
worldsheet. In section three we describe the actions we get on the worldsheet from the gravity backgrounds
discussed in section two. In section four we discuss in more detail some particular backgrounds. First we discuss
the background leading to the $N=2$ sine Gordon model on the worldsheet and the associated duality to the $CP^1$
model. We then
  discuss what happens if we have an
$A_N$ singularity transverse to a pp-wave and we resolve it.

\newsec{Supersymmetric supergravity solutions of pp wave  type }

We consider
type IIB supergravity solutions with a nonzero 5-form field
strength. They have a covariantly constant null killing vector,
${\partial \over \partial x^-}$, which
also leaves $F_5$ invariant and it is such that it gives zero when
contracted with $F_5$.

More explicitly,
the form of the solutions we consider is
\eqn\ansatz{
\eqalign{ &ds^2=-2dx^+dx^-+H(x^i)(dx^+)^2+ds_8^2 \cr &F_5=dx^+\wedge \varphi_4(x^i) }
}
where $x^i$ are the 8 transverse coordinates, $F_5$
is the self-dual RR field strength. We limit ourselves to
solutions which are also independent of $x^+$. We consider constant
dilaton and set all other fields to zero.
The transverse metric can be  curved. Note  that the background is
such that we can scale down  $H$ and $\varphi$ by  performing
 a boost in the $x^\pm$ directions.\foot{So
 the background is not boost invariant in the $x^\pm$ directions.}.
 This property under boost transformations implies that we can assign an
``order'' to each field according to how they change under boosts. The four-form $\varphi$ is of first order while
$H$ is of second order.
 This means that the transverse space with zero RR five-form should be a solution
of the equations of motion by itself, since it is of zeroth order.

In order to clarify a bit the discussion we will first consider the
simpler case when  the transverse space is flat and then the slightly
more complicated  case of a curved transverse space.

\subsec{Flat transverse space}

The equations of motion of type IIB supergravity imply that \ansatz\
obeys
\eqn\eoms{ \nabla^2 H = -32|\varphi|^2 \;\; ; \;\;\;\; ~~~~~
  *_{10}F_5 = F_5
} where $|\varphi|^2 = {1\over 4!}\varphi_{\mu\nu\rho\delta} \varphi^{\mu\nu\rho\delta}$ , and $\nabla^2$ is the
laplacian in the transverse 8-dimensional space. In our conventions\foot{Our conventions and notations are
summarized in Appendix A.}, the  self-duality of $F_5$ implies that $\varphi$ is anti-self-dual in the
8-dimensional space,
 so that  $*\varphi = -\varphi$ and $d\varphi=0$. \foot{
A $*$ with no subindex will always refer to the 8 dimensional
space.}

In addition we will now require the solution to preserve some
supersymmetries. Supersymmetries in type IIB supergravity are
generated by a chiral spinor $\epsilon$ with 16 complex
components. We find it convenient to separate it into two
components according to their $SO(8)$ chiralities
\eqn\chiraldecompsition{ \epsilon = -{1\over 2}\Gamma_+\Gamma_-
\epsilon - {1\over 2}\Gamma_-\Gamma_+\epsilon \equiv \epsilon_+
+\epsilon_-~. } $\epsilon_+ $ has positive $SO(1,1)$ and $SO(8)$
chiralities, and is annihilated by $\Gamma_{\hat{+}}$. We will
find, roughly speaking (i.e. to lowest order in $\varphi_4$), that
$\epsilon_+$ is related to the supersymmetries that are preserved
by a configuration with nonzero $p_-$ and are linearly realized on
the
 light cone action. These  anti-commute to the lightcone Hamiltonian, plus possibly some rotations.
On the other hand the supersymmetries generated by $\epsilon_-$ are non-linearly realized on the worldsheet and
 imply that some particular fermions are free on the worldsheet. For reasons that will become clear later we
are especially interested in supersymmetries that are linearly realized on the worldsheet so we are interested
in spinors such that only $\epsilon_+$ is nonzero to first order.

Setting to zero the supersymmetry variations we obtain the following equation
\eqn\susygeneq{
0=D_{M}\epsilon =(\nabla_{M} +{i\over 2}\sf\Gamma_{M})\epsilon ~,
}
which leads to
\eqn\epplus{\eqalign{
 &\partial_-\epsilon_+ =
\partial_\mu \epsilon_+= \partial_+\epsilon_+=0 \cr &
\partial_-\epsilon_- =0 \;\;\;\; ; \;\;\;\;
\partial_\mu\epsilon_- = {i\over
2}\Gamma_-\sv\Gamma_\mu\epsilon_+ \;\;\;\; ; \;\;\;\; (i\partial_+ -\sv)\epsilon_- = {i\over 4}\Gamma_-\sp
H\epsilon_+ } } where $\sv\equiv { 1 \over 4 !}
\Gamma^{\mu\nu\rho\delta}\varphi_{\mu\nu\rho\delta}$. These equations imply that
$\epsilon_+$ must be a constant spinor and they determine the first and higher order parts of $\epsilon_-$ in
terms of $\epsilon_+$. These solutions with nonzero zeroth order  $\epsilon_+$ determine the linearly realized
supersymmetries of the light cone action. In addition to these we might have solutions of \epplus\ with
$\epsilon_+ =0 $. We obviously have 16 solutions of this type if $\varphi$ is a constant form, but when $\varphi$
is not constant we will generically have no solutions of this type (below we will make a precise statement). Note
that only solutions of this second type can be $x^+$ dependent. Note also that
 if $\epsilon=\epsilon_++\epsilon_-$ is a solution, then so is
$\hat{\epsilon}=\epsilon_+^*-\epsilon_-^*(-x^+)$.

When we  attempt to solve  the equation for  $\epsilon_-$ in terms of $\epsilon_+$ we  find some integrability
conditions. First, integrability of the
 $\partial_\mu \epsilon_-$ equations places a constraint
on the allowed 4-forms. Then the $(i\partial_+-\sv)\epsilon_-$ equation gives
 further consistency conditions on $\epsilon_-$ and determines $H$
in terms of $\varphi_4$. In Appendix B we show these computations in detail. Below we will just state the form of
the most general solutions with $(2,2)$ and $(1,1)$ supersymmetry. We did not explore the subset of $(2,2)$
solutions which actually have more $\epsilon_+$-type supersymmetries.

It is convenient to choose complex coordinates for the transverse space, $z^1,...,z^4$. The anti-self-dual 4-forms
$\varphi_{\mu\nu\rho\delta}$ written in complex coordinates can be split into 2 kinds - those having two
holomorphic and two anti holomorphic
 indices - the (2,2) forms (of which there are 15) and those
having one holomorphic and three antiholomorphic indices
 and their complex conjugates - the (1,3) and (3,1) forms
(of which there are 10+10).
We denote the (1,3)
forms by the shorter notation
\eqn\shortnot{
\varphi_{mn} \equiv {1\over
3!}\varphi_{m\overline{ijk}}\epsilon^{\overline{ijkn}}g_{n\bar{n}}
}
Anti-self duality of $\varphi$ implies that $\varphi_{mn}$ is symmetric.

It can be shown that one can write
the anti-selfdual  $(2,2)$ forms in terms of
$\varphi_{i\bar j}$ defined as
\eqn\shorttwo{
 2 \varphi_{l \bar m} = g^{s \bar s} \varphi_{l \bar m s \bar s}  ~,
}
where the reality and  self duality condition imply that
 $\varphi_{l \bar m} $ is a hermitian and traceless matrix (which could,
in principle, be a function of the coordinates). We also define
the lowest weight spinor state $|0\rangle $ which is annihilated
by $\Gamma_{\hat{+}}$ and $\Gamma^i$ where $i$ runs over the four
holomorphic indices. We begin by describing the solutions with an
$\epsilon_+$ which at zeroth order is proportional to $|0\rangle$
and its complex conjugate. We later describe solutions with
$\epsilon_+ =0$.

\item{CASE (1)}  (2,2) {\it supersymmetry or more}

The solution is parameterized by a holomorphic function $W$. In this case the $\varphi_{l \bar m}$ are  constants
and given in terms of a
 traceless hermitian 4x4 matrix. $W$ and $\varphi_{l \bar m}$ should also obey
\eqn\eqforc{
\partial_n[\varphi_{j}^{~~k}{z^j}\partial_kW]=0
}
where we raised the index of $\varphi_{j \bar k}$ using the flat transverse space  metric.
The metric and the  4-form are given by
\eqn\solcaseI{\eqalign{
 ds^2 &= -2dx^+dx^- -32(|\partial_kW|^2 +
|\varphi_{j\bar{k}}z^j|^2)(dx^+)^2 +dz^i\overline{dz^i}
\cr \varphi_{mn} &=\partial_m\partial_nW  ~,~~~~~
\varphi_{\bar m {\bar n} } =  \partial_{\bar m}\partial_{\bar n}
 \overline W  ~,~~~~~~~ \varphi_{l \bar m} = { \rm constants}
}}
The expressions for the Killing spinors can be found in appendix B.

One can, of course, look at the simpler cases where either $W=0$ or
$\varphi_{l \bar m} =0$. It is interesting to note that if $\varphi_{l \bar m}$ is nonzero the superalgebra
has a central charge term proportional to the $U(1)$ symmetry generated by the holomorphic Killing vector
$z^l \varphi_{l \bar m}  \partial/\partial z^m $ and its complex conjugate.

\item{ CASE (2)}  (1,1) {\it supersymmetry}

These solutions are parameterized by a real harmonic
function $U$.
 However this time there are only
 2 Killing spinors.
 The solution is \eqn\caseII{\eqalign{ ds^2 &= -2dx^+dx^- -32( |\partial_k U |^2 ) (dx^+)^2
+dz^id\overline{z^i} \cr \varphi_{mn} &=
\partial_m\partial_n U  \;\;\; ; \;\;\;
\varphi_{\bar m \bar n} =
\partial_{\bar m} \partial_{\bar n} U  \;\;\; ; \;\;\;
\varphi_{l\bar{m}} = \partial_l\partial_{\bar m} U
 }}
The expressions for the Killing spinors can be found in appendix B.

\subsec{ The homogenous solution for $\epsilon_-$}

The homogenous equations for $\epsilon_-^{hom}$ are
\eqn\homeq{
\partial_-\epsilon_-^{hom}=\partial_j\epsilon^{hom}_-=
\overline{\partial_j}\epsilon^{hom}_-=(i\partial_+-\sv)\epsilon^{hom}_-=0 }
and are solved by \eqn\xpl{ \epsilon_-^{hom} (x^+) = e^{-i\sv x^+}\eta_0 }
 where
$\eta_0$  is a  constant spinor. \homeq\ implies that $\sv$ and $\eta_0$
should be such that after multiplying $(\sv)^n \eta_0$ (for
$n=1,2,...$) we still have spinors that are constant in the transverse space and independent of $x^+$. So we get
the spinors $ \eta_0 , \sv \eta_0 , \cdots (\sv)^{n -1} \eta_0$ which are linearly independent and $n \leq 16$.
These solutions of \homeq\ are associated to
 free fermions on the string worldsheet in light cone gauge. In fact the
last equality in \homeq\ is the equation of motion for a zero
momentum mode on the string worldsheet. If we diagonalize the
matrix $\sv$ in the subspace of solutions we see clearly that each
pair of solutions gives rise to a free fermion on the
worldsheet\foot{ The solutions come in pairs. If the eigenvalue of
the matrix $\sv$ is nonzero this follows by considering the
complex conjugate equation. If the eigenvalue is zero then we can
multiply the solution by any complex number so that we have two
real solutions.}. The fermion is free but it can be massless or
massive depending on the eigenvalue of the matrix $\sv$~ on it.
The sixteen supersymmetries of $\epsilon_-$ type that arise in the
usual quadratic
 plane waves discussed in  \Blau\  arise
because all fermions are free. In a general interacting case all fermions will be interacting and there will be no
supersymmetries of this type. If, in addition, we have worldsheet supersymmetry in lightcone gauge, as in the
cases we are analyzing, each free fermion has a free boson partner and these two together  decouple from the rest
of the worldsheet  theory. So the structure is clear, we have as many free bosons and fermions as there are
$\epsilon_-$ supersymmetries. In the N=(2,2) case these supersymmetries come in groups of four, one per complex
field that appears at most  quadratically in the superpotential.

\subsec{ Curved transverse space}

When the transverse space is curved, the ansatz (2.1) is a solution of IIB supergravity iff it satisfies the
equations of motion \eqn\eomscur{\eqalign{ \nabla^2H = -32|\varphi|^2 \;\;\; ; \;\;\; *_8 \varphi &= -\varphi
\;\;\; ; \;\;\; d\varphi=0 \cr R_{\mu\nu} &=0 }} where $\nabla^2$
 is the laplacian  in the transverse curved space, and
$R_{\mu\nu}$ is the Ricci tensor of the transverse space \foot{we
use (+,-) and greek letters to denote curved indices , and (v,u)
and roman letters to denote flat indices. All notations and
conventions we use for curved space are summarized in Appendix A.
}.

The supersymmetry equations for the curved case are \eqn\eppluscrv{\eqalign{ &\partial_-\epsilon_+ = \nabla_{\mu}
\epsilon_+= \partial_+\epsilon_+=0 \cr &\partial_-\epsilon_- =0 \;\;\;\; ; \;\;\;\; \nabla_{\mu}\epsilon_- =
{i\over 2}\Gamma_u\sv\Gamma_{\mu}\epsilon_+ \;\;\;\; ; \;\;\;\; (i\partial_+ -\sv)\epsilon_- = {i\over
4}\Gamma_u\sp H\epsilon_+ }} These are exactly the same equations as in the flat case \epplus , with the
transverse derivatives replaced by  covariant derivatives. We will now state what the general solutions are and we refer
the interested reader to appendix B for the derivation. The first point to note is that to zeroth order
 the supersymmetry equations for the
transverse manifold imply that the transverse space is a special
holonomy space. If we demand  (2,2) supersymmetries on the
worldsheet it can only be a Calabi-Yau space ($G_2$ and $Spin(7)$
could also be studied but we do not do that here). For this reason
it is still convenient to choose complex coordinates and we denote
by $|0\rangle$ the covariantly constant spinor on the Calabi-Yau
manifold that is annihilated by $\Gamma_v$ and $\Gamma^{\mu}$
where $\mu$ runs over the four holomorphic indices.  We will also
use the short notation \shortnot\ for
 the (1,3) forms.
We first focus on the supersymmetries that are linearly realized on the
worldsheet in lightcone gauge and later we explain what happens with
the homogeneous solutions for $\epsilon_-$.

\item{ CASE (1)}  (2,2) {\it supersymmetry or more}

In this case the solution is parameterized by a holomorphic function $W$, and a real  Killing potential $U$ from
which we can define the Killing vectors $V_\mu = i \partial_{\mu} U $, $ V_{\bar \mu} = - i \partial_{\bar \mu} U
$. The Killing vector should be
 holomorphic (i.e. $V^\mu$ is holomorphic and $V^{\bar \mu}$ is antiholomorphic). The
following conditions should also hold
\eqn\trac{ \nabla_{\mu}V^{\mu} = 0}
\eqn\symofw{
\partial_{\nu}[V^{\tau}\nabla_\tau W ] =0
}
 The supergravity solution is
\eqn\curi{\eqalign{ ds^2 &= -2dx^-dx^+ -32(|dW|^2 +|V|^2 )(dx^+)^2
+2 g_{\mu \bar \nu}dz^{\mu}d{\bar z}^{\bar \nu}
 \cr \varphi_{\mu\nu} &= \nabla_{\mu}\nabla_{\nu}W ~,~~~~~~~~~~~~
 \varphi_{\bar \mu\bar \nu} = \nabla_{\bar \mu}\nabla_{\bar \nu} \bar W\cr \varphi_{\bar \mu \nu} & =
\nabla_{\bar \mu} \nabla_\nu U
}}
 where $|dW|^2 \equiv
g^{\mu\bar{\nu}}\nabla_{\mu}W\overline{\nabla_{\nu}W}$, and $|V|^2 \equiv g_{\mu\bar{\nu}}V^{\mu}V^{\bar{\nu}}$.
The expressions for the Killing spinors can be found in appendix C.

Here too, one can look at the simpler cases where either $W=0$ or $V^{\mu}=0$.
Note that if the transverse space is compact there is no  non-constant holomorphic function.
In order to have interesting solutions we need the transverse space to be non-compact.

\item{ CASE (2)} (1,1) {\it supersymmetry }

The (1,1) supersymmetry solutions are parameterized by a real harmonic function $U$.
The metric, 4-form and the 2 Killing spinors are given by
\eqn\curiv{\eqalign{ ds^2 &= -2dx^-dx^+ -32( |\nabla U|)^2 (dx^+)^2 +g_{\mu\bar{\nu}}z^{\mu}\overline{z^{\nu}} \cr
\varphi_{\mu\nu} &= \nabla_{\mu}\nabla_{\nu} U  \;\;\; ; \;\;\; \varphi_{\bar \mu \bar \nu} = \nabla_{\bar \mu}
\nabla_{\bar \nu} U  \;\;\; ; \;\;\; \varphi_{\mu\bar{\nu}} = \nabla_\mu\nabla_{\bar \nu} U
 }}
Note that the (2,2) part of the 4-form (whose components are $\varphi_{\lambda\bar{\sigma}\mu\bar{\nu}}$ ) is
therefore  \eqn\vphitwo{ \varphi = (\nabla_{\mu}\nabla_{\bar \nu}U dz^{\mu}d\overline{z^{\nu}})\wedge J}
where $J$ is the Kahler form, which obeys $dJ=0$ (so  that $\varphi_{\mu\bar{\nu}}={1\over 2}
g^{\lambda\bar{\sigma}}\varphi_{\lambda\bar{\sigma}\mu\bar{\nu}} =\nabla_{\mu}\nabla_{\bar{\nu}}U$).

\subsec{ The homogenous solution for $\epsilon_-$}

The homogenous equations for $\epsilon_-^{hom}$ in a curved background are \eqn\homeqcur{
\partial_-\epsilon_-^{hom}=\nabla_j\epsilon_-^{hom}=\overline{\nabla_j}\epsilon_-^{hom}=(i\partial_+-\sv)\epsilon_-^{hom}=0 } There is a
solution  \eqn\xplcur{ \epsilon_-^{hom}(x^+)=e^{-i\sv x^+}\eta_0} with $\eta_0$ a covariantly constant spinor and
all of $(\sv)^n\eta_0$ ($n=1,2,...$) covariantly constant.

The discussion follows exactly the one we had for the flat case, where we argued that each pair of solutions for
\homeqcur\
 gives rise to a free (massive or massless) fermion on the string worldsheet in light cone gauge. Due to
supersymmetry each such fermion has a free boson partner, and they both decouple from the rest of the
 worldsheet theory.

\newsec{ The worldsheet actions}

In the last section we have listed all the supersymmetric solutions of the pp-wave form. In this section we write
the action describing a string propagating in these backgrounds.
We choose light cone gauge by setting $x^+ = \tau$, where $\tau$ is worldsheet time. Though the standard
procedure we then find that $p_-$ is conserved, etc.\foot{
Our notation with a lower  index for $p_\pm$ seems to be contrary to standard practice in the literature. While
in Minkowski space it does not matter where we put the index, it actually
does matter  where we put it when $g_{++}$ is
nonzero. (Some papers have chosen the unreasonable convention of raising  the indices using the
flat Minkowski metric...). In our conventions for the metric (where $g_{-+} =-1$)
we find that $p_- \leq 0$ for particles propagating
to the future. }
 In light cone gauge, only killing spinors
which are not annihilated by $\Gamma^+$ survive as linearly realized supersymmetries on the worldsheet. These are
the $\epsilon_+$ part of the killing spinor. Since we focused on solutions that preserved some supersymmetries of
this type, we will have a supersymmetric action on the worldsheet. Thanks to these supersymmetries we do not need
to work too much to find the action, since its form is dictated by supersymmetry.

{\bf \item{(2,2)}  Supersymmetric solutions}

We know that if all RR fields are set to zero, the action reduces to the usual $(2,2)$ non-linear sigma model
which can be written in terms of the K\"{a}hler potential. By turning  on $(1,3)$ and $(3,1)$ forms we can add an
{\it arbitrary} superpotential so that the action in superfield form becomes \eqn\LGsup{\eqalign{ S =& {1\over
{4\pi\alpha'}}\int d\tau \int_0^{2\pi\alpha' |p_-|}d\sigma (L_K + L_W ) ~, \cr
 &L_K + L_W =
\int d^4\theta K(\Phi^i,\bar{\Phi}^i) +{1\over 2}(\int d^2\theta W(\Phi^i)+c.c.)
}}
 where $\Phi^i = Z^i
+\sqrt{2}\theta^L\psi^i_L +\sqrt{2}\theta^R\psi^i_R +2\theta^L\theta^RF^i
 +... $. {}From this we can  find the component action by integrating out
$\theta$ \wbag .
 Note that \LGsup\ contains
Yukawa interactions given in terms of $\sv$,
 a bosonic potential proportional to $H$ \ansatz , as well as four
fermion couplings  which follow from supersymmetry. If the transverse space is flat, there are no four fermion
couplings, and the action could also be read from \Metsaev . The fermions appearing in \LGsup\ are related to the
Green-Schwarz fermions as follows. The G-S fermions are SO(8) spinors with negative chirality (in our
conventions). Once we choose complex coordinates we have an SU(4) subgroup of SO(8) which preserves the complex
structure. Under this subgroup $8_- \to {\bf 4} + {\bf \bar 4}$, these are the spinors with vector index. More
explicitly, let us denote by $\eta_0$ a covariantly constant spinor annihilated by all $\Gamma_{\bar i}$. We then
write a general negative chirality SO(8) spinor as $ S = \psi^i \Gamma_{i} \eta_0 + \psi^{\bar i} \Gamma_{\bar i}
\eta_0^* $. This defines the worldsheet spinors $\psi^i$, $\psi^{\bar i}$.

It can be checked that the $(3,1)$ and $(1,3)$ forms induce couplings of the type $\psi^i_L \psi_R^j$ as
implied by the action \LGsup . It can also be seen that the (2,2) forms induce couplings of the type
$\psi^i_L \psi_R^{\bar j}$. These couplings are not present in \LGsup .
Nevertheless, it was shown in
 \WitBag , \Freed , \Gates, and reviewed in \LG, that if the target space has a holomorphic isometry, i.e. a
holomorphic killing vector field $V^i$  ($\nabla_iV_{\bar{j}}+\nabla_{\bar{j}}V_i=0$), then this isometry can be
gauged to give a vector multiplet (consisting of a complex scalar, two conjugate dirac fermions and a vector
field). Then by taking the weak coupling limit and then freezing the vector and fermions at zero and the scalar at
a constant value, one can obtain a (2,2) supersymmetric lagrangian. The extra terms in the Lagrangian that arise
in this way are \eqn\LVec{ L_V = -g_{i\bar{j}}|m|^2V^i\overline{V^j} -{i\over 2}(g_{i\bar{i}}\partial_jV^i -
g_{j\bar{j}}\partial_{\bar{i}}\overline{V^j})(m\overline{\psi^i_R}\psi^j_L +\bar{m}\overline{\psi^i_L}\psi^j_R)
~.}

Note that in our case, we cannot obtain any such holomorphic Killing vector - we have the extra requirement
(coming from the self-duality of $F_5$) that $\nabla_{\mu}V^{\mu}=0$. It might be possible that including more
background fields, such a three form RR field strength,  we get a more general Lagrangian.

In the simple case where the transverse space is flat, we have a holomorphic killing vector
$V_{\bar{j}}=ic_{i\bar{j}}z^i$ , for a hermitian constant matrix $c_{i\bar{j}}$, and $\nabla_{\mu}V^{\mu}=0$
translates into the tracelessness of $c_{i\bar{j}}$.

The combined action coming from $L_K+L_W+L_V$ is supersymmetric iff $V^{\mu}\nabla_{\mu}W$ is constant
\Freed . This matches nicely with the condition \symofw .

{\bf \item{(1,1)} Supersymmetric solutions }

A general (1,1) supersymmetric sigma-model is of the form \eqn\nonelagr{ S={1\over {4\pi\alpha'}}\int d\tau
\int_0^{2\pi\alpha' |p_-|}d\sigma d^2 \theta ( g_{\mu\nu} D_L\phi^\mu D_R \phi^\nu +  U(\phi) ) } where $\phi^\mu$
are $N=1$ superfields.
The
superpotential $U(\phi)$ is not as general as it could be in an arbitrary $N=1$ theory, since it needs to be a harmonic
function. This condition also follows from conformal invariance in
the Berkovits formulation \nbjm .
 Of course if we view the $N=(2,2)$ solution as an $N=(1,1)$ theory then the corresponding $N=1$
superpotential is harmonic due to the stricter
 constraints that both the superpotential and
Killing potential of the $N=2$ theory have to obey.

\newsec{ Some examples }

In this section we discuss
some general features of the models and describe in more detail some examples.

\subsec{ RG flow}

The light cone worldsheet theory is a theory with a mass scale. So these theories behave quite non-trivially under
RG transformations. This mass scale on the worldsheet is basically set by $p_-$. More precisely the important
dimensionless parameter is  $\alpha' |p_-| \mu$ where $\mu$ is the coefficient in front of the
superpotential $ W = \mu f( z/\l_s)$ where $f$ is a dimensionless function. This dimensionless parameter is
the product of the mass scale on the worldsheet and the size of the worldsheet cylinder.
A physical spacetime question, like the spectrum of the theory, depends non-trivially on this dimensionless
parameter. We see that performing a scale transformation on the worldsheet is related to performing a boost
in the $x^+, ~x^-$ coordinates. For low values of $|p_-|$ we are exploring the UV of the  worldsheet
theory while for
large values we explore the IR. As usual we have a UV/IR relation between worldsheet and target space
scales.  Note that in many situations, most notably the
$c<1$ string theories, one can start with a non-conformal theory and ``dress'' it with the Liouville mode so
that the total theory is a critical string theory. In those cases the RG flow in the original massive theory
becomes related to a change in position
 along the Liouville direction. Notice that this case has a different
 character since
an RG transformation is related to a change in {\it velocity} of the motion in the $x^+, ~ x^-$ direction.
In other words in one case we have that an RG transformation is a {\it translation}
 in the Liouville direction whereas in
our case it is a  {\it boost} in the $x^+, ~x^-$ directions.
The worldsheet will generically have periodic boundary conditions for the fermions since they are Green-Schwarz
fermions. The number of zero energy (zero $p_+$) supersymmetric ground states can be computed by the standard
index arguments. These will be BPS states in the spacetime theory.

It is interesting to note that we can choose a superpotential that has no supersymmetric vacua. In this
case  we do not have a supersymmetric vacuum on the worldsheet which means that the
corresponding state in the spacetime theory  is not BPS when $p_-$ is non-zero. Supersymmetry breaking on the
worldsheet should not be confused with spacetime supersymmetry breaking.

\subsec{Solitons}

One feature of our models is that they contain solitons on the worldsheet. The worldsheet is compact and has a
size proportional to $ |p_-| \alpha'$. If $|p_-|$ is large we will be able to trust soliton computations which are
done in an infinite line. Note that when the string is propagating with fixed value of $p_-$ it feels a
gravitational force that pulls it to the regions where $-g_{++}$ is a  minimum.
  A soliton on the worldsheet going between
these minima corresponds to a string that goes between the two positions where $- g_{++}$
 has a minimum in target space.
For example, we can choose a superpotential which is a function of only one variable $W(z_1)$. In this case the
three other complex fields on the worldsheet are massless and free. If we solve $\partial_{z_1} W =0$ we will
obtain the values of $z_1$ corresponding to supersymmetric vacua in the field theory. The gravitational force will
be directed towards these points in spacetime. We can have string configurations that interpolate between these
different points. However, as we are discussing closed strings of finite length (i.e. we impose periodic boundary
conditions on the worldsheet), these configurations will not be topologically stable, unless there are
identifications in the transverse space. We will discuss below a case with identifications in the transverse
space.

\subsec {Integrable theories}

It is possible to choose the superpotential in such a way that we get an integrable model on the worldsheet. We
can then rely on the large literature on integrable models to derive properties of the worldsheet theory. Of
course the most interesting regime is when the worldsheet theory is strongly coupled, since in this case we do not
have any other simple method to derive the spectrum. Our above derivation of the lightcone worldsheet lagrangian
is only valid for weak coupling, since we used the supergravity approximation. It is nevertheless possible to show
that in the case of flat transverse space these are good string solutions by using one of Berkovits' formalisms
\berkovits\  \nbjm . We now  take a flat transverse space and we explore the physics that results from adding a
superpotential of the form $ W(z^1) = \lambda \cos \omega z^1  $. This gives the $N=2$ supersymmetric sine Gordon
theory. More explicitly the full background is \eqn\ppssg{\eqalign{ ds^2 &= -2dx^+dx^- -| {\lambda  \omega}\sin
\omega z^1|^2 (dx^+)^2 + dz^id\overline{z^i} =\cr &= -2dx^+dx^- -{1\over 2}|{\lambda \omega}|^2[\cosh (2\omega
x^5) - \cos (2\omega x^1)](dx^+)^2 +dx^idx^i \cr F_5 &= dx^+\wedge \varphi_4 \;\;\;\; ; \;\;\;\;
\varphi_4={\lambda \omega^2\over 32} \cos (\omega z^1) dz^1\wedge \overline{dz^2}\wedge \overline{dz^3}\wedge
\overline{dz^4}+c.c. }} where $z^1 = x^1 + i x^5$. The sine Gordon model is conventionally written in terms of
canonically normalized fields $\phi = z/\sqrt{ 2 \pi \alpha'}$ and the parameter $\beta $ is defined by writing
the superpotential as $ W =  \mu \cos \beta \phi$ (where $\mu$, which is proportional to $\lambda$, has dimensions
of mass). This implies that $ w = { \beta \over \sqrt{ 2 \pi \alpha'} }$. At this point we could consider two
models, one where $x^1$ is non-compact or another were $x^1 $ is compact. Below we will be interested in the model
where $x^1 \cong x^1 + 2 \pi/\omega$. This model is such that we have two distinct supersymmetric vacua, $x^1 = 0,
{\pi \over \omega}$ (and also $x^5 =0$). When we consider this sine Gordon model on an infinite spatial line (and
time) one can compute exactly its S-matrix \Kob . It was found that the S-matrix is the product of the S-matrices
for two theories, one is an integrable version of the $N=2$ minimal models and the other is the S-matrix of the
bosonic sine Gordon theory. The $N=2$ minimal model is the one with $Z_2$ global symmetry. The spectrum contains a
kink and anti-kink together with some breathers of masses \eqn\masses{ M_n = 2 m_s \sin ( {n \pi \over 2 \gamma})
~,~~~~~~~~\gamma = { 8 \pi \over \beta^2} } where $n=1,...,N$ and  $N =[\gamma]$ is the number of breathers. $m_s$
in \masses\ is the mass of a soliton which is proportional to  $\mu$. In order to find the spectrum of states in
string theory we need to find the spectrum of the sine Gordon theory on a circle. If the size of the circle is
very large, which corresponds to large $|p_-|$, we can use the Bethe ansatz to obtain an approximate answer for
the spectrum. The corresponding expression is expected to be correct up to exponentially small corrections in the
size of the circle (or $e^{ - |p_-| \mu \alpha'}$). Some exact results for the spectrum on the cylinder for a
simple integrable model were  obtained in \zamlu , but as far as we know the spectrum for the $N=2$ sine Gordon on
the cylinder is not known.

Note that the limit $\beta \to 0$ corresponds to the semiclassical limit of the sine Gordon model. In this limit
the period of the sine is much longer than $\alpha'$. This means that the background $F$ field involves large
length scales. In this limit there is a large number of breathers. The lowest lying breather is the basic
perturbative massive field in the theory and the lowest lying ones can be thought of as bound states of these. On
the other hand the limit of large  $\beta$ corresponds to the  quantum regime of the sine Gordon model. Note that
for $\gamma <1$ there are no breathers, we only have the kinks and anti-kinks. When $\beta$ is large the radius of
the $x^1$ circle in string units is small so that one would attempt to do a T-duality on this circle. Since the
background
 fields depends explicitly on $x^1$ this is not a straightforward
T-duality. Fortunately the
necessary transformation is the  mirror symmetry transformation discussed in \refs{\LG,\hk},
which gives a sausage model. In fact
this relation was conjectured first in \fintr , by studying the S-matrices and
it is a
close relative of \zamsausage .  The radius of the sausage is proportional to $\beta$. More precisely it is
$\tilde R = \alpha'  \omega $.
We can see that in the limit that the RR fields are small, which is the UV of the
worldsheet theory then in the original picture we have a cylinder with a gravitational potential that confines
the strings to the region near the origin of the non-compact direction along  the cylinder.
 In the T-dual picture we have a cylinder of the T-dual radius  near the central region of the original cylinder,
but
 the compact circle of the cylinder shrinks as we move away from the center so that we form a sausage.
The sausage model is again not conformal invariant so that the geometry of the sausage depends on the scale. As we
go to the UV of the field theory on the worldsheet the sausage becomes longer and longer as $\log(E)$, where $E$
is the energy in question. Of course such a model contains a mass scale which is basically set by $|p_-|$. When we
go to the IR the sausage model develops a mass gap and there are only a few massive excitations. We conclude that
we have a background which is such that if we explore it with strings that have low values of $|p_-|$ we see it as
being very large, while if we explore it with strings with higher values of $|p_-|$ it appears smaller. A natural
question that arises is whether this background is a solution of the supergravity equations. For large values of
$\tilde R$, which means large values of $\beta$, the curvature of the sigma model is small so one would expect it
to be a solution of supergravity. In particular the $\beta =\infty$ limit is the $SU(2)$ symmetric round $CP^1$
model \FendleyB .
 On the other hand, one could make an
argument that this background cannot be a simple supergravity solution, at least within the context of a simple
light cone reduction.  The reason is  the complicated way
in which the scale of the model determines the geometry. When we go to light cone the scale that appears in the
light cone theory is related to $\partial X^+$. If this scale appears quadratically or linearly in the
lightcone action it is very simple to find the particular supergravity fields that give rise to the light cone
gauge model, quadratic appearances of $\partial X^+$ are related to $g_{++}$ and linear appearances of
$\partial X^+$ are related to fields with one $+$ index, such as $F_{+ \cdots}$. In the round $CP^1$ model
the scale is appearing schematically as
\eqn\action{ S \sim \int \log(E/|p_-|) \partial \theta \partial \theta \sim
 \int \log (E/ |\partial X^+| ) \partial \theta \partial \theta
} in the action, where the last term is {\it very} schematic.
This suggests that the background leading to
this $CP^1$ model contains excited massive string modes. In fact, if we treat the
RR field as
a small perturbation (which is correct if we are near the center of the cylinder and at small
$|p_-|$) we can see that a T-duality in the the $x^1$ direction would transform the momentum mode
of $F_5$ into a winding mode (with winding number two). This is somewhat reminiscent of the
description of the cigar used in \kkk , though in that case one could view the background as a gravity
solution. Another related, but distinct, way in which a massive theory as the $CP^1$ model could arise in
string theory was presented in \tseytlinmass . In that case the RG direction was precisely $x^+$ and the
metric was $x^+$ dependent.

All that we said here about the sine Gordon model can be generalized to affine Toda theories (with rank smaller
than five) \FendleyB .  The mirror symmetry transformation in this case will produce a deformed $CP^N$ model \hk .

\subsec{ Resolving $A_N$ singularities}

In this section we will consider deformations of $A_N$ singularities in the presence of RR fields.\foot{ This
problem was also considered in \Floratos , where some singular solutions were described. Here we construct
non-singular solutions. } We can start with the maximally supersymmetric plane wave of IIB theory which has a
field strength of the form $\varphi_{1234} = - \varphi_{5678}$=constant  and all other components equal to zero.
We can form complex coordinates $z^j = x^j + ix^{j+4}$. Then we see that this background corresponds to a
background with zero (2,2) forms and a superpotential of the form $W = \mu \sum_{i=1}^4 (z^i)^2 $. We can consider now
the $R^4$ space spanned by the coordinates $1256$ and replace it by an $A_N$ singularity.
This background still preserves half the
supersymmetries. Let us start discussing first the case of an $A_1$ singularity.
We see that we can replace the $A_1$ singularity by the Eguchi Hanson space, which is a Ricci
flat K\"ahler (actually hyperK\"ahler) manifold. When the RR fields are zero this solution
 preserves the same number of
supersymmetries as the $A_1$ singularity. They preserve 8 supersymmetries that are linearly realized on the
worldsheet, which is actually a (4,4) theory. We also have 8 other supersymmetries that are non-linearly realized
and which are associated to the four real coordinates spanned by $z^3,z^4$ which are free on the worldsheet.

Another interesting situation
 to consider is an $A_1$
singularity involving the first four coordinates $1234$. In this
case, in order to find a supersymmetric deformation, it is convenient to group the coordinates into complex
coordinates as $z^1 = x^1 + i x^2$, $z^2 = x^3 + i x^4$, etc. Then the maximally supersymmetric solution can be
thought of as a solution with $ W=0$ and only (2,2) forms with Killing potential $U = \mu (|z_1|^2 + |z_2|^2
-|z_3|^2 - |z_4|^2)$. We can still resolve the $A_1$ singularity by replacing
it by an Eguchi-Hanson space. In this case the solution will be of the type described in section 2.3. The Killing
potential is
\eqn\kptn{
 U = \mu [\sqrt{1+{a^4\over \rho^4}}(|z^1|^2 + |z^2|^2) -(|z^3|^2 + |z^4|^2)] =
\mu [r^2-(|z^3|^2 + |z^4|^2)]~,
}
 where $\rho^2 \equiv |z^1|^2+|z^2|^2$ , $r^4\equiv \rho^4+a^4$, and $a$ is the Eguchi-Hanson
resolution parameter. The derivatives of $U$ form a holomorphic Killing vector $V^{\nu} = -ig^{\nu{\bar \nu}}
\partial_{\bar{\nu}}U = -i\mu (z^1,z^2,-z^3,-z^4) $ and the (2,2) forms are given by $\varphi_{\nu\bar \sigma} =
\nabla_{\nu}\nabla_{\bar{\sigma}}U$. One can see that the solution actually has (4,4) supersymmetry since one
can redefine the coordinates $z_{3,4} \to \bar z_{3,4}$ and construct new Killing spinors of the type we constructed
above. Furthermore if we view the theory as an $N=1$ theory the superpotential we get in both cases is the same, so that
we have twice the number of supersymmetries.
Potentials for $(4,4)$ two dimensional theories were considered in \refs{\jor,\Freed}.
In conclusion, we have a (4,4) theory on the lightcone worldsheet.
Of course we also have another 8 supersymmetries of the $\epsilon_-$ type that are due to the fact that the
coordinates $z_3,~z_4$ are free.

Above we discussed supersymmetric deformations of the $A_1$ singularity. There are also non-supersymmetric
deformations, which we can describe most easily by writing the Eguchi Hanson metric in real coordinates
  \eqn\eh{
ds^2  = {dr^2 \over (1 -{a^4 \over r^4})} + {r^2\over 4} (d\theta^2 + \sin^2\theta d\phi^2) + {r^2\over 4} (1
-{a^4 \over r^4})( d\psi + \cos\theta d\phi)^2 } where the angles take values in $\theta \in [0,\pi )$ ; $\phi ,\psi
\in [0,2\pi )$. Then we can choose the four form to be proportional to the volume element, and the metric
component $g_{++} = - \mu r^2$ looks the same as what it was for the original $A_1$ singularity. This solution is
{\it not} supersymmetric. It differs from the supersymmetric solution by some terms which are localized near the
singularity. We can view the non-supersymmetric solution as the supersymmetric one plus some normalizable modes
that live near the singularity. These are  normalizable modes of the four form potential. From the point of view
of the worldvolume theory on the $A_1$ singularity, these are  the modes that gives rise to the self dual tensor in
six dimensions. Indeed one can check that the difference between the 5-form field-strengths of the two solutions
is $\Delta F_5 \sim h_3 \wedge l_2$, where $h_3 = h_{+ ij}$ is an anti-self dual tensor on the six directions
corresponding to the worldvolume of the resolved $A_1$ singularity (i.e. directions $+-5678$)
 and $l_2$ is the unique normalizable anti-self dual
two form on the Eguchi Hanson space,  $l_2 =  {1\over r^2}[{2\over r}dr\wedge (d\psi +\cos\theta
d\phi)-\sin\theta d\theta\wedge d\phi]$.

The solution considered in \Floratos ~is equal to the non-supersymmetric solution described above,
 up to the addition of a harmonic function to $g_{++}$ ,which is singular at $r=0$.
For any of the solutions described in this paper, we can add a singular harmonic function of the transverse
coordinates  to $g_{++}$. We can  think of them as describing the metric generated by massless particles
with worldlines along $x^-$.

Of course all that we said above can be extended to $A_{N-1}$
singularities by replacing the Eguchi-Hanson instanton
by the geometry of the resolved ALE space.
These $A_{N-1}$ singularities arise as Penrose limits of $AdS_5 \times S^5/Z_{N}$, it would be nice to know
if in this case we can also resolve the singularity in a smooth fashion.
In the case of $(AdS_3 \times S^3)/Z_N$ we know that we can smooth out the singularity in simple way \martinec .

\newsec{Open problems}

It would be nice to obtain some more exact results for strings propagating on these backgrounds and explore
further what they teach us about strings on non-trivial RR backgrounds. In particular, it would be nice to
understand further the target space interpretation of the Sine-Gordon model at large $\beta$. It is clear that we
can add D-branes to these backgrounds. These D-branes are expected to
 be supersymmetric if they sit on holomorphic submanifolds where $W$ is
a constant or on mid-dimensional lagrangian submanifolds such that the
image in the $W$ plane is constant
 \hvopen \foot{
We thank M. Gaberdiel for correcting us on this point.}.
 One could  explore a more general ansatz where we also have a non-zero three form RR
field strength. An interesting question is  if there are any supersymmetric deformations of $A_N$ singularities
when they are embedded in $AdS_5 \times S^5$. Of course, it would be nice to find a holographic dual for these
backgrounds.

{\bf Acknowledgments}

We would like to thank D. Berenstein, N. Berkovits,
R. Gopakumar, K. Hori, K. Intrilligator,  M. Moriconi, H. Nastase,
  N.~Seiberg and  J.~Sonnenschein
for useful discussions.

This  research was supported in part by DOE grant
DE-FG02-90ER40542.

\vfill

\appendix{A}{Conventions and notations, and the supersymmetry equations}

{\it Flat transverse space}

We use conventions where $x^\pm \equiv {1\over \sqrt{2}}(x^0\pm x^9)$ and $\epsilon_{+-12345678}=+1$.
$F_5=dx^+\wedge \varphi_4$. Since $F_5$ is self-dual and closed $\varphi_4$
is
anti-self-dual in the transverse 8-dimensions and closed.
For the metric \ansatz\ with flat transverse space
we choose the vielbiens  as $\theta^{\hat{i}}=dx^i, ~  \theta^{\hat{+}}=dx^+ , ~
\theta^{\hat{-}}=dx^--{1\over 2}Hdx^+$.
The corresponding connections all vanish except $\omega^{\hat{-}i}=-\omega^{i\hat{-}}=-{1\over
2}\partial_iHdx^{{+}}$.
The covariant derivatives acting on spinors are $\nabla_-=\partial_- , ~ \nabla_i=\partial_i ~,
\nabla_+=\partial_+-{1\over 4}\partial_iH\Gamma_-\Gamma_i$.
And the terms involving  $F_5$ in the IIB covariant derivative are
$\sf \Gamma_- = \Gamma^+\sv\Gamma_-=0$, $\sf \Gamma_j = -\Gamma_-\sv\Gamma_j$, $\sf\Gamma_+ =
-\Gamma_-\sv\Gamma_+$.
The chirality matrix is  $\Gamma_{11}=-\Gamma^{01...89} =  {1\over 2} [\Gamma^+,\Gamma^-] \Gamma^{1...8}$.
The IIB spinor is a 16-component complex chiral spinor satisfying $\Gamma_{11}\epsilon=+\epsilon$.
Since $\varphi_4$ is anti-self-dual in 8-dimensions, acting on a chiral spinor $\sf\Gamma_+\epsilon =
 2\sv\epsilon$.
Using all the above, the susy equations $D_{\mu}\epsilon \equiv (\nabla_\mu-{i\over 2}\sf\Gamma_\mu)\epsilon=0$
take the form \foot{To relate these conventions to the ones in  Blau, Figueroa et al  \Blau\  take their
conventions, replace their $x^\pm$ with $x^{0,9}$ according to $x^\pm = {1\over \sqrt{2}}[x^9 \pm x^0]$. take $x^0
\to -x^0$ then flip one of the coordinates, say $x^1 \to -x^1$, and then replace back with chiral coordinates
$x^\pm_{here} = {1\over \sqrt{2}}[x^0 \pm x^9]$.} \eqn\susyflat{\partial_-\epsilon=0 ~~;~~ \partial_+\epsilon-
({1\over 4}\Gamma_-\sp H-i\sv)\epsilon =0 ~~;~~
\partial_j\epsilon-{i\over 2}\Gamma_-\sv\Gamma_j\epsilon=0 }
We would find it easier to work in complex coordinates, so we split the transverse space ($x^1,...,x^8$) to 4
complex coordinates $z^j =  x^j+ix^{j+4}$.
In complex coordinates, the susy equations \susyflat ~are
\eqn\susycmplx{\eqalign{ &\partial_-\epsilon=0 \cr &\partial_+\epsilon - ({1\over
4}\Gamma_-\bar{\Gamma}\cdot\bar{\partial}H+{1\over 4}\Gamma_-\Gamma\cdot\partial H -i\sv)\epsilon =0 \cr
&\partial_j\epsilon-{i\over 2}\Gamma_-\sv\Gamma_j\epsilon=0 \;\;\; ; \;\;\; \bar{\partial}_j\epsilon-{i\over
2}\Gamma_-\sv\overline{\Gamma_j}\epsilon=0 }}
Let us classify the a.s.d 4-forms according to their holomorphicity properties. Denoting by
(p,q) the number of holomorphic and anti-holomorphic indices in $\varphi_{abcd}$ ($p+q=4$), there are 10
(1,3)-forms, 10 (3,1)-forms, and 15 (2,2)-forms, giving a total of 35 a.s.d. 4-forms. The (2,2) forms are of the
form $\varphi_{i\bar{i}j\bar{k}}$ and $\varphi_{i\bar{i}j\bar{j}}$ (no sum), and a.s.d implies that
$\varphi_{1\bar{1}2\bar{2}}=-\varphi_{3\bar{3}4\bar{4}}$ etc. and
$\varphi_{1\bar{1}2\bar{3}}=\varphi_{4\bar{4}2\bar{3}}$ etc. The (3,1) and (1,3) forms are of the form $
\varphi_{\bar{i}jkl} , \varphi_{\bar{i}ijk}, \varphi_{i\overline{jkl}} , \varphi_{i\overline{ijk}}$, and a.s.d.
relates $\varphi_{1\overline{123}}=-\varphi_{4\overline{423}}$ , $\varphi_{\bar{1}123}=-\varphi_{\bar{4}423}$ etc.
The closed condition
 relates the (2,2) to the (1,3), (3,1) components. The reality condition on $\varphi$ implies that
$\varphi_{i\overline{jkl}}=\varphi_{\bar{i}jkl}^*$, $\varphi_{i\bar{j}k\bar{l}}=\varphi_{j\bar{i}l\bar{k}}^*$.

Going  back to the susy equations \susycmplx , we separate $\epsilon$ into two components of different transverse
chiralities $ \epsilon = -{1\over 2}\Gamma_+\Gamma_-\epsilon - {1\over 2}\Gamma_-\Gamma_+\epsilon \equiv
\epsilon_++ \epsilon_-$. Since $\epsilon$ has a positive $\Gamma_{11}$ chirality, $\epsilon_+$ has positive
$SO(1,1)$ and $SO(8)$ chiralities, and $\epsilon_-$ has both negative. The susy equations for $\epsilon_+$ are
$\partial_-\epsilon_+ =
\partial_j\epsilon_+=\overline{\partial_j}\epsilon_+=(\partial_++i\sv)\epsilon_+=0$.
As $\varphi$ has negative $SO(8)$ chirality, automatically, $\sv\epsilon_+=0$ and we conclude that $\epsilon_+$
must be a constant spinor.
The susy equations for $\epsilon_-$ are
 \eqn\epplusA{\eqalign{ \partial_-\epsilon_-=0 ~~~~~~&;~~~~~~ (i\partial_+
-\sv)\epsilon_- = {i\over 4}\Gamma_-\sp H\epsilon_+ \cr \partial_j\epsilon_- = {i\over
2}\Gamma_-\sv\Gamma_j\epsilon_+ ~~~~~~&;~~~~~~
 {\partial_{\bar j}}\epsilon_- = {i\over 2}\Gamma_-\sv{\Gamma_{\bar j}}\epsilon_+
\cr} } In order to solve the susy equations explicitly, it is
convenient to introduce a Fock space notation.  The vacuum
$|0\rangle$ is defined to be the spinor annihilated by
$\Gamma_{\hat{+}}$ and by all $\Gamma^i$ (where $i$ is a
holomorphic index) . We also define the operators $b^{ i}
=\Gamma^i = g^{i \bar j} \Gamma_{\bar j}$, $b^{+ \bar i} =
\Gamma^{\bar i}$.
 Note that in this normalization $\{b^i,b^{+ \bar j}\}=2g^{i \bar j}$, where $g^{i \bar  j}$ is the inverse
of the Kahler metric. This is not  the usual normalization of annihilation and creation operators.
We denote $ \varphi_{mn} \equiv {1\over
3!}\varphi_{m\overline{ijk}}\epsilon^{\overline{ijkn}}g_{n\bar{n}}, ~
\varphi_{\overline{mn}}\equiv (\varphi_{mn})^*$
 (so that
e.g. $\varphi_{24}=\varphi_{2\overline{123}}$ , $\varphi_{21} = -\varphi_{2\overline{234}}$). Anti-self-duality implies that
$\varphi_{mn}=\varphi_{nm}$ , $\varphi_{\overline{mn}}=\varphi_{\overline{nm}}$. We also use the notation
$2\varphi_{m\bar{n}}\equiv g^{s \bar s}\varphi_{s\bar{s}m\bar{n}}$, and denote by
$  \widetilde{b}^k | 0 \rangle \equiv
b^k { 1 \over 4 \ 4!} \epsilon_{\bar i \bar j \bar k \bar l}
(b^{+ \bar i } b^{+ \bar j}b^{+ \bar k} b^{+ \bar l})|0\rangle$ a 'hole' creation operator acting on the vacuum.
The slashed four-form
acts on the Fock space states as
\eqn\svonsp{\sv b^{+ \bar m}
|0\rangle= 4 [\varphi^{\bar m}_{~~ n} \widetilde{b}^{ n}-
\varphi^{\bar m}_{~~ \bar n}b^{ + \bar n}]|0\rangle \;\;\;\; ; \;\;\;\; \sv \widetilde{b}^m
|0\rangle=4[\varphi^m_{~~ \bar n} b^{+\bar n}-\varphi^m_{~~  n} \widetilde{b}^n]|0\rangle
}
where  we have raised the indices of $\varphi_{ab}$ using the metric.
We parameterize $\epsilon_\mp$ in this Fock space
\eqn\epscomps{\epsilon_- =
\Gamma_-[\beta_{\bar k} b^{+\bar k} +\delta_k \widetilde{b}^k]|0\rangle ~~~~;~~~~ \epsilon_+=[\alpha+{1\over
2}\gamma_{\bar p \bar q}b^{+\bar p} b^{+\bar q} +\zeta { \epsilon_{\bar i \bar j \bar k \bar l}
(b^{+ \bar i } b^{+ \bar j}b^{+ \bar k} b^{+ \bar l}) \over 4 \ 4! } ]|0\rangle }
$\alpha,\gamma_{\overline{pq}}, \zeta$ are complex constants, and $\beta_{\bar m} ,\delta_k$ are
complex functions
of $z^i,\overline{z^i}$.
By an appropriate SO(8) rotation we will see that we can set $\gamma_{\bar p \bar q} $ to zero in our solutions.
So from now on we set it to zero.
Using \svonsp\ one can check that
 \eqn\susyone{\eqalign{
\sv\epsilon_- &= -4\Gamma_- [
\beta_{\bar m}\varphi^{\bar m}_{~~ \bar n} - \delta_ m\varphi^{ m }_{~~ \bar n}]
b^{+ \bar n} |0\rangle
+ 4 \Gamma_-
[\beta_{\bar m}\varphi^{\bar m}_{~~  n}
- \delta_m \varphi^{ m}_{~~  n} ]
\widetilde{b}^n
|0\rangle
\cr
 \snabla H\epsilon_+ &=
\alpha \partial_{\bar j} H
b^{+ \bar j}|0\rangle +
\zeta \partial_j H
\widetilde{b}^{ j} |0\rangle }}
The susy equations become the following equations for
$\alpha,\beta_{\bar m},
\delta_m ,\zeta$
\eqn\susyb{\eqalign{
& 4 (\beta_{\bar m} \varphi^{\bar m}_{~~ n}- \delta_{ m} \varphi^m_{~~ n}) = -{i\over 4}
\zeta\partial_n H
+i\partial_+\delta_n \cr &
4 (-\beta_{\bar m} \varphi^{\bar m}_{~~ \bar{n}} +\delta_m\varphi^m_{~~ \bar n } ) = - {i\over 4}
\alpha\partial_{\bar n}H
+i\partial_+\beta_{\bar{n}}
\cr
&\partial_j\beta_{\bar{k}} = -2i \alpha\varphi_{j\bar{k}}
   ~~;~~
\partial_{\bar j}\beta_{\bar{k}} =
2i  \zeta\varphi_{\overline{jk}}
\cr & \partial_{\bar j} \delta_k =
-2i \zeta\varphi_{k\bar{j}}
  ~~~~~;~~~
\partial_j\delta_k = 2 i
\alpha\varphi_{jk}
 }}

{\it Curved transverse space}

Starting from the metric
\eqn\met{ ds^2 = -2dx^+dx^- +H(x^\rho)(dx^+)^2+g_{\mu\nu}(x^\rho)dx^\mu dx^\nu }
the
nonzero connections for this metric are
$ \Gamma^-_{++} = -{1\over 2}\partial_+H \; ; \; \Gamma^-_{+\mu}=-{1\over 2}\partial_\mu H \; ; \; \Gamma^\mu_{++}
= -{1\over 2}g^{\mu\nu}\partial_\nu H \; ; \; \Gamma^\mu_{\nu\rho} = \gamma^\mu_{\nu\rho} $ , where
$\gamma^\mu_{\nu\rho}$ are the connections on the 8-dimensional manifold. The only components of the Ricci tensor
which do not vanish are $R_{++}$ and $R_{\mu\nu}$ which are given by $R_{++} = -{1\over 2}\nabla^2 H \; ; \;
R_{\mu\nu} = r_{\mu\nu}$, where $r_{\mu\nu}$ is the ricci tensor for the 8-dimensional metric. The Ricci scalar is
the same as that of the 8-dimensional metric $R=r$. The Einstein equations are then  $r_{\mu\nu}=0$  and
$\nabla^2 H= -32|\varphi|^2$ , where $|\varphi|^2\equiv {1\over
4!}\varphi_{\mu\nu\rho\delta}\varphi^{\mu\nu\rho\delta}$.
We also introduce the corresponding flat indices $a=(v,u,i,j,...)$ and the coframe $\theta^v=dx^+ ;\;\;\; \theta^u
= dx^--{1\over 2}Hdx^+ ;\;\;\;\; \theta^i_\mu dx^\mu$, such that $ds^2 =-2\theta^v\theta^u +\sum_i
\theta^i\theta^i$. The connections are determined by the no torsion condition and their nonzero components are
$\Omega^u_{\;\;i} = -{1\over 2}\theta^{\mu}_i\partial_\mu Hdx^+ , ~ \Omega^i_{\;\;j} =
\omega_{\mu\;j}^i(x^\rho)dx^\mu  $, where $\omega^i_{\;\;j}(x^\rho)$ are the connections on the 8-dimensional
manifold, satisfying $d\theta^i+\omega^i_j\wedge\theta^j=0$. The covariant derivatives $\nabla_M = \partial_M +
{1\over 2}\Omega_M^{ab}\Gamma_{ab}$ are given by \eqn\covd{ \nabla_- = \partial_- \;\;;\;\; \nabla_\mu =
\partial_\mu + {1\over 2}\omega_\mu^{ij}\Gamma_{ij} \;\;;\;\; \nabla_+ = \partial_+ -{1\over
4}\theta^{i\mu}\partial_\mu H\Gamma_{ui} }
And the susy equations $0=D_M\epsilon=(\nabla_M +{i\over 2}\sf\Gamma_M)\epsilon$ are therefore
\eqn\susycurv{\eqalign{ &\partial_-\epsilon=0 ~~~~;~~~~ \partial_+\epsilon- {1\over 4}\Gamma_u\sp H \epsilon
+i\sv\epsilon =0 \cr &[\partial_\mu +{1\over 2}\omega_\mu^{ij}\Gamma_{ij}]\epsilon-{i\over
2}\Gamma_u\sv\Gamma_\mu\epsilon=0 }}
The above equations are exactly the ones we had before for the flat case \susyflat, the only difference being
trading the regular derivative in the 8-dim space with a covariant derivative. also we recall that the Einstein equations
 imply
$g_{\mu\nu}$ is Ricci flat. Let us now try to solve these
equations, similarly to what we did in the flat case. Again we
change to complex coordinates, and separate
$\epsilon=\epsilon_-+\epsilon_+$. As before, we get that
$\epsilon_+$ must be a covariantly constant spinor, i.e.
\foot{From here on $\nabla_{\mu}$ denotes a covariant derivative
in the 8-dimensional transverse space.}
$\partial_-\epsilon_+=\partial_+\epsilon_+=\nabla_\mu\epsilon_+=\overline{\nabla_\mu}\epsilon_+=0$.
The equations for $\epsilon_-$ are \eqn\susycurv{\eqalign{
\partial_-\epsilon_- &=0 \cr \nabla_{\mu}\epsilon_-={i\over
2}\Gamma_u\sv\Gamma_\mu\epsilon_+  \;\;\;\; &; \;\;\;\;
\overline{\nabla_{\mu}}\epsilon_-={i\over
2}\Gamma_u\sv\overline{\Gamma_\mu}\epsilon_+ \cr
(i\partial_+-\sv)\epsilon_- &={i\over 4}\Gamma_u\sp H \epsilon_+
}} As in the flat case, we again use the notation
$\varphi_{\mu\nu}$, and introduce the Fock space $|0\rangle$ which
is annihilated by $\Gamma_v$ and by all $\Gamma^{\mu}$ ($\mu$ a
holomorphic curved index), and is a covariantly constant
spinor\foot{ As the manifold is a CY, there is a covariantly
constant spinor
 $\psi_0 = |0\rangle$. The spinor $|0\rangle$ is actually constant. In fact
the Killing
spinor equation is  $
\partial_{\mu} |0\rangle +{1\over
2}\omega^{i\bar{j}}_{\mu}\Gamma_{i\bar{j}} |0 \rangle =0$.
The term  $\Gamma_{i\bar{j}}|0\rangle $  is proportional to $g_{i\bar{j}}$
and therefore to the trace of the spin-connection, which on a CY can be chosen to be zero \GSW. },
 and the operators $b^{\bar{\mu}+}\equiv
\overline{\Gamma^\mu}=\theta^{\bar{\mu}}_{\bar{i}}\overline{\Gamma^i}\;\;$
; $\;\;b^\mu \equiv \Gamma^\mu = \theta^\mu_i\Gamma^i\;\;$;
$\;\;\{b^\mu,b^{\bar{\nu}+}\}=2g^{\mu\bar{\nu}}$. From now on we
can define the ``hole'' operator $ \widetilde{b}^\mu$ as we did in
the flat space case. Similarly we can define $\beta_{\bar \mu}$,
$\delta_\mu$, $\alpha$ and $\zeta$ as in \epscomps . We can
similarly derive equations \svonsp \susyone\ and finally \susyb ,
where all that we would need to do is to replace the ordinary
derivative with covariant derivatives for the transverse indices.

\appendix{B}{ Derivation of the flat space supersymmetric solutions}

We have seen that $\epsilon_+$ should be a constant. As the transverse space is $R^8$ we can always do an SO(8)
transformation which sets $\gamma_{\bar p \bar q} =0$ in \epscomps ,
 but we will be unable to distinguish solutions with (2,2) susy from solution with more susy.
We also set all $x^+$ dependence to zero, because, as discussed before, this part could always be added as a
solution to the homogenous equations.
Integrability of the $\partial_j\delta_k$ and $\partial_{\bar j}\beta_{\bar{k}}$ in \susyb\ then assures (as
$\alpha,\zeta$ are not both zero) that the (1,3) and (3,1)-forms make a closed form by themselves.
Using the fact that the (1,3) and (3,1) parts of $\varphi$ are separately anti-self-dual and closed, we can show that
$\varphi_{ij}$ satisfies $\varphi_{ij}=\varphi_{ji}$ from anti-self-duality,
$\partial_{[i}\varphi_{j]m} = g^{\bar k k } \partial_{\bar k} \varphi_{kj}=0$ from
closedness,
for all $i,j,m$.
These imply that $\varphi_{ij}=\partial_i\partial_jW$ where $W$ is a harmonic function . Similarly, as
$\varphi_{m\bar{n}}$ must be hermitian and  closed by themselves, they must be of the form
$\varphi_{m\bar{n}}=\partial_m\partial_{\bar{n}}\cal{U}$ where $\cal{U}$ is a real harmonic function. The
equations \susyb (with no $x^+$ dependence) become
 \eqn\susyd{\eqalign{ (\beta^m \partial_m\partial_n W
-\delta^{\bar m} \partial_{\bar{m}} \partial_n{\cal U}) &= -{i\over 16} \zeta\partial_n H \cr
-(\beta^m \partial_n\partial_{\bar{n}}{\cal U} -\delta^{ \bar m} \partial_{\bar m}\partial_{\bar n}
\overline{W}) &= - {i\over 16}
\alpha \partial_{\bar n} H  \cr
\partial_j\beta_{\bar{k}} = -2i\alpha\partial_j\partial_{\bar{k}}{\cal U} \;\; &; \;\; \partial_{\bar j}\beta_{\bar{k}} =
2i\zeta \partial_{\bar j}\partial_{\bar k}\overline{W}\cr \partial_{\bar j} \delta_k = -2i
\zeta\partial_k\partial_{\bar{j}}{\cal U} \;\; &; \;\; \partial_j\delta_k =  2i
\alpha\partial_j\partial_kW\cr }}
Integrability of the equations implies that
\eqn\slkjf{(|\zeta|^2-|\alpha|^2)\partial_{\bar{j}}\partial_m\partial_kW
=(|\zeta|^2-|\alpha|^2)\partial_m\partial_{\bar{j}}\partial_k{\cal U}
=(|\zeta|^2-|\alpha|^2)\partial_{\bar{m}}\partial_{\bar{j}}\partial_k{\cal U} = 0~,}
for all $m, \bar{m},\bar{j},k$. This can be satisfied in one of the following two cases

{\bf (i)} $|\alpha|\neq |\zeta|$, $W$ is holomorphic and harmonic, and
$\varphi_{j\bar{k}}=\partial_j\partial_{\bar{k}}\cal{U}$ is a 4x4 hermitian traceless matrix of {\it constants}.
In that case we can solve the $\partial_j$ and $\overline{\partial_j}$ equations to get \foot{There is no need to
add integration constants to $\beta_{\bar{k}} , \delta_k$, as such terms can be set to zero by a redefinition of
$dW$ by a constant shift, and a redefinition of $z^j$ by a constant shift.}
 \eqn\betadeltaii{ \beta_{\bar{k}} =
-2i[\alpha \varphi_{j\bar{k}}z^j -\zeta\overline{\partial_kW}] \;\; ; \;\; \delta_k = -2i [\zeta
\varphi_{k\bar{j}}\overline{z^j}-\alpha\partial_kW ]}
 Then plugging these back into the first two equations in \susyd, and taking into account the fact that $H$
is real, we get the {\it consistency condition}
$\partial_n[\varphi_{j\bar{k}}z^j\partial_kW]=0$, and the
expression for  $ H = -32
(|\partial_kW|^2+|\varphi_{j\bar{k}}z^j|^2)$\foot{Here too there
is no need to add an integration constant to $H$, as such a
constant can be set to zero, shifting $x^-$ by a constant times
$x^+$.}. This is the solution with (2,2) supersymmetries, or more,
that we have in \solcaseI . Plugging \betadeltaii\ in  \epscomps\
we get the explicit expression for the four Killing spinors, which
are parameterized by the two complex numbers $\alpha, ~\zeta$.

{\bf (ii)}  $|\alpha|=|\zeta|$.   Now we have that  for all $i,j,\bar{k}$
$\partial_i\partial_j\partial_{\bar{k}}[{\cal U}+{\alpha\over \zeta} W]=0$. Without loss of generality , we
choose the constant phase ${\alpha \over \zeta}=-1$.\foot{ This amounts to redefining the complex coordinates by
a constant phase.} Then one can define $U$ a real harmonic function such that
$\partial_j\partial_kU=\partial_j\partial_k W$ and
$\partial_j\partial_{\bar{k}}U=\partial_j\partial_{\bar{k}}{\cal U}$, so the four-form is given by the second
derivatives of $U$
 \eqn\svone{\varphi_{ij}=\partial_i\partial_j U \;\;\;\; ; \;\;\;\;
\varphi_{\bar{i}j}=\partial_{\bar{i}}\partial_jU \;\;\;\; ; \;\;\;\;
\varphi_{\overline{ij}}=\partial_{\bar{i}}\partial_{\bar{j}}U ~.}
 Solving the $\partial_j$ and
$\overline{\partial_j}$ equations gives
 \eqn\betadeltai{ \beta_{\bar{k}} = 2i\zeta\partial_{\bar{k}}U \;\; ; \;\;
\delta_k = -2i\zeta
\partial_kU }
Then plugging these into the first two equations gives two
identical equations for $H$ , which are solved by
$H=-32|\partial_k U|^2$. These are the (1,1) supersymmetric
solutions we have in \caseII . Plugging \betadeltai\ into
\epscomps\ we get the explicit expression for the two Killing
spinors that are parameterized by one complex number, $\alpha = -
\zeta$.

\appendix{C}{ Derivation of the curved space supersymmetric solutions}

Here too we set $\gamma_{\overline{\mu\nu}}=0$ . This way we would still find all solutions with at least (1,1)
supersymmetry, but would not be able to distinguish solutions with (2,2) supersymmetry from solutions with more
supersymmetry. Note that if the transverse space has precisely SU(4) holonomy then the Killing spinor
has $\gamma_{\bar \mu \bar \nu} =0$.
 We also take as in the flat case, $\beta_{\bar{\nu}} , \delta_{\nu}$ to be independent of $x^+$
(the $x^+$ dependent part would be dealt with as part of the solution to the homogenous equations for
$\epsilon_-$).
Then the equations that we get from \susyb\ by replacing ordinary derivatives by
covariant derivatives becomes.
\eqn\susybcurv{\eqalign{
& 4 (\beta_{\bar \mu} \varphi^{\bar \mu}_{~~ \nu}- \delta_{ \mu} \varphi^\mu_{~~ \nu}) = -{i\over 4}
\zeta\partial_\nu H
\cr &
4 (-\beta_{\bar \mu} \varphi^{\bar \mu}_{~~ \bar{\nu}} +\delta_\mu\varphi^\mu_{~~ \bar \nu } ) = - {i\over 4}
\alpha\partial_{\bar \nu}H
\cr
&\nabla_\mu\beta_{\bar{\nu}} = -2i \alpha\varphi_{\mu\bar{\nu}}
   ~~;~~
\nabla_{\bar \mu}\beta_{\bar{\nu}} =
2i  \zeta\varphi_{\overline{\mu \nu}}
\cr & \nabla_{\bar \mu} \delta_\nu =
-2i \zeta\varphi_{\nu\bar{\mu}}
  ~~~~~;~~~
\nabla_\mu\delta_\nu = 2 i
\alpha\varphi_{\mu\nu}
 }}
The integrability conditions for $\nabla\delta$ and $\overline{\nabla}\beta$ imply that
$\nabla_{[\rho}\varphi_{\mu]\nu}=0$ (i.e. the (1,3) and (3,1) forms are closed by themselves). Thus
$\varphi_{\mu\nu} = \nabla_\mu\nabla_\nu W$ for some harmonic function $W$. The (2,2) forms therefore should be
closed by themselves, and together with anti-self-duality
 they must satisfy $ \varphi_{\mu \bar \nu}
= \nabla_{\mu}\nabla_{\bar{\nu}}{\cal U} = \nabla_{\bar{\nu}}\nabla_{\mu}{\cal U}$ for some real harmonic function
${\cal U}$. Plugging these back to the equations \susybcurv\ , we get
 \eqn\susycurvnogamb{\eqalign{
\nabla_\mu\beta_{\bar{\nu}} = -2i\alpha \nabla_{\mu}\nabla_{\bar{\nu}}{\cal U}
   \;\;\;\;\;\;\;\; ; \;\;\;\;\;\;\;\; &\overline{\nabla_\mu}\delta_{\nu} = -2i\zeta \nabla_{\bar{\mu}}\nabla_{\nu}{\cal
   U}
 \cr \overline{\nabla_\mu}\beta_{\bar{\nu}} = 2i\zeta \nabla_{\bar \mu}\nabla_{\bar \nu}\overline{W}
   \;\;\;\;\;\;\;\;\;\; ; \;\;\;\;\;\;\;\; &\nabla_{\mu}\delta_{\nu}
 = {2i}\alpha\nabla_{\mu}\nabla_{\nu}W \cr  -
[\beta^\rho
\nabla_{\rho}\nabla_{\bar{\nu}}{\cal U}
-\delta^{\bar \tau}
\nabla_{\bar \tau}\nabla_{\bar \nu} \overline{W} ] &= -{i\over 16} \alpha
\partial_{\bar \nu}\overline{H}
\cr
 [\beta^\tau \nabla_{\tau}\nabla_{\mu}W
- \delta^{\bar \tau }\nabla_{\mu}\nabla_{\bar{\tau}}{\cal U}] &= -{i\over 16} \zeta
\partial_\mu H }}
We can immediately solve the two equations in the first line to get
$\beta_{\bar{\nu}}=-2i\alpha\nabla_{\bar{\nu}}{\cal U}+f_{\bar{\nu}}(\bar{z}), ~
\delta_{\nu}=-2i\zeta\nabla_{\nu}{\cal U}+g_{\nu}(z)$ for some antiholomorphic and holomorphic one-forms
$f_{\bar{\nu}}(\bar{z}) , g_{\nu}(z)$ respectively. Then we can plug these back into the two equations in the
second line, and get the constraints
\eqn\cons{ \nabla_{\mu}[\nabla_{\nu}(\zeta{\cal U}+ \alpha
W)+{i\over 2} g_{\nu}(z)]= \nabla_{\mu}
[\nabla_{\nu}(\alpha^*{\cal U}+\zeta^* W)-{i \over 2} f_{\bar{\nu}}^*(z)]= 0 .} These can be
solved in one of two ways.

{\bf (i)} $|\alpha| \neq |\zeta|$

Then we can define a new real harmonic function $U$ related to ${\cal U}$ through $f_{\bar{\nu}} , g_{\nu} $
\foot{ The relation is $U \equiv {\cal U} +{(i\zeta^*\int g_{\nu}dz^{\nu} +c.c.) \over
2[|\zeta|^2-|\alpha|^2]}+ {(i\alpha\int f_{\nu}dz^{\nu}+c.c.) \over 2[|\zeta|^2-|\alpha|^2]}$.} such that
$\nabla_{\mu}\nabla_{\bar{\nu}}U=\nabla_{\mu}\nabla_{\bar{\nu}}{\cal U}$ , and by \cons\
$\nabla_{\mu}\nabla_{\nu}U=0$. Note that $U$ is a Killing potential,  if we define a vector
$V_{\mu}=i\nabla_{\mu}U$ then $\nabla_{\bar{\mu}}V^{\nu}=\nabla_{\mu}V^{\bar{\nu}}=0$ and
$\nabla_{\mu}V_{\bar{\nu}}+\nabla_{\bar{\nu}}V_{\mu}=0$. This means that $V^{\mu}$ is a {\it holomorphic Killing
vector}. Additionally, as $U$ is a harmonic function, the Killing vector also satisfies $\nabla_{\mu}V^{\mu}=0$.
By \cons  , one also finds that $\nabla_{\mu}\nabla_{\nu}W$ is holomorphic. Since $W$ appears in the susy
equations only under two holomorphic covariant derivatives, we can take $W$ to be holomorphic. One can now solve
the first four equations in \susycurvnogamb\ to get \foot{ We did not include integration constants in $\beta_{\nu}
, \delta_{\nu}$ as these can always be set to zero be a redefinition of the potentials.}
\eqn\betdel{
\beta_{\bar{\nu}} = 2i[i\alpha V_{\bar{\nu}} +\zeta \overline{\nabla_{\nu}W}] \;\;\;\; ; \;\;\;\;
\delta_{\nu}={2 i}[-i\zeta V_{\nu}+\alpha\nabla_{\nu}W]~,
}
 where $\varphi_{\mu\nu}=\nabla_{\mu}\nabla_{\nu}W$ and $ \varphi_{\mu \bar \nu} =
\nabla_{\mu}\nabla_{\bar{\nu}}U $. Then plugging these into the
last two equations in \susycurvnogamb , and using the fact $H$ is
real, we get one constraint on $W$ and $V^{\mu}$ and one equation
for $H$. The constraint is
$\partial_{\nu}[V^{\tau}\nabla_{\tau}W]=0$,  and the equation for
$H$ yields $ H = -32(|dW|^2 +|V|^2)$, where $|dW|^2 \equiv
g^{\mu\bar{\nu}}\nabla_{\mu}W\overline{\nabla_{\nu}W}$ and $|V|^2
\equiv g_{\mu\bar{\nu}}V^{\mu}V^{\bar{\nu}}$. This is the (2,2)
supersymmetric solution we have in \curi . Inserting \betdel\ into
\epscomps\ we get the explicit expression for the four preserved
Killing vectors parameterized by $\alpha, ~\zeta$.

{\bf (ii)} $|\alpha|=|\zeta|$.  We can define a real harmonic function $U$ such that
$\nabla_{\mu}\nabla_{\nu}U = \nabla_{\mu}\nabla_{\nu}W$ and $\nabla_{\mu}\nabla_{\bar{\nu}}U =
\nabla_{\mu}\nabla_{\bar{\nu}} {\cal U}$, so that  $\varphi_{\mu\nu}=\nabla_{\mu}\nabla_{\nu}U$ ,
$\varphi_{\mu\bar{\nu}}=\nabla_{\mu}\nabla_{\bar{\nu}}U$ ,
$\varphi_{\overline{\mu\nu}}=\nabla_{\bar{\mu}}\nabla_{\bar{\nu}}U$. Then solving for $\beta_{\bar{\nu}}$ and
$\delta_{\nu}$ , one gets \eqn\betdelb{ \beta_{\bar{\nu}} = 2i\zeta \nabla_{\bar{\nu}}U \;\;\;\; ; \;\;\;\;
\delta_{\nu} = -2i\zeta\nabla_{\nu}U} Plugging these back into the last two equations \susycurvnogamb , one gets
the same equation for $H$, whose solution is $ H = -32|dU|^2$. These are the (1,1) supersymmetric solutions we
have in \curiv . Again we can insert \betdelb\ in \epscomps\ to get the explicit expression for the Killing spinors.

\listrefs

\bye